\documentclass[sigconf,nonacm]{acmart}

\usepackage{booktabs}
\usepackage{subcaption}
\usepackage{dblfloatfix}
\usepackage{balance}

\begin{document}

\title{Deep Joint Embeddings of Context and Content\\for Recommendation}

\author{Miklas S. Kristoffersen}
\authornote{The work was done in part while the author was visiting BBC R\&D in London, UK. His work is supported by the Innovation Fund Denmark (IFD) under File No. 5189-00009B.\\ Presented at CARS 2.0: Context-Aware Recommender Systems Workshop  - in conjunction with the 13th ACM Conference on Recommender Systems (RecSys'19).}
\affiliation{%
  \institution{Bang \& Olufsen A/S}
  \institution{BBC R\&D}
  \institution{Aalborg University}}
\email{mko@bang-olufsen.dk}

\author{Jacob L. Wieland}
\affiliation{%
  \institution{BBC}
  \city{London}
  \country{United Kingdom}}
\email{jacob.wieland@bbc.co.uk}

\author{Sven E. Shepstone}
\affiliation{%
  \institution{Bang \& Olufsen A/S}
  \city{Struer}
  \country{Denmark}}
\email{ssh@bang-olufsen.dk}

\author{Zheng-Hua Tan}
\affiliation{%
  \institution{Aalborg University}
  \city{Aalborg}
  \country{Denmark}}
\email{zt@es.aau.dk}

\author{Vinoba Vinayagamoorthy}
\affiliation{%
  \institution{BBC R\&D}
  \city{London}
  \country{United Kingdom}}
\email{vinoba.vinayagamoorthy@bbc.co.uk}

\fancyhead{}

\begin{abstract}
    This paper proposes a deep learning-based method for learning joint context-content embeddings (JCCE) with a view to context-aware recommendations, and demonstrate its application in the television domain.
    JCCE builds on recent progress within latent representations for recommendation and deep metric learning.
    The model effectively groups viewing situations and associated consumed content, based on supervision from 2.7 million viewing events.
    Experiments confirm the recommendation ability of JCCE, achieving improvements when compared to state-of-the-art methods.
    Furthermore, the approach shows meaningful structures in the learned representations that can be used to gain valuable insights of underlying factors in the relationship between contextual settings and content properties.
\end{abstract}


\maketitle

\section{Introduction}
Recommender systems (RS) have evolved rapidly in recent years thanks to widespread attention in both academia and industry~\cite{Adomavicius2005a,Shi2014,Veras2015}.
One of the major developments has been the introduction of deep learning-based methods, which have demonstrated superior performance in numerous applications~\cite{Zhang2019}. 
These deep methods have proven to be able to effectively capture nonlinear and nontrivial relationships in the interactions between users and items.
Beyond users and items, context-aware RS (CARS) have seen new exciting opportunities.
The ability of deep neural networks to learn underlying explanatory factors and low-dimensional representations from sparse input data with a large number of attributes, is key to achievements of recent latent CARS, e.g. as demonstrated in an unsupervised setting in~\cite{Unger2016}.
In combination with the rich descriptive information of user-item interactions that is available in several modern real-world applications, these methods provide a promising way forward in the endeavor to understand users and provide them with context-aware recommendations.

\begin{figure}[tb]%
    \centering%
    \vspace{1em}%
    \includegraphics[width=0.9\columnwidth]{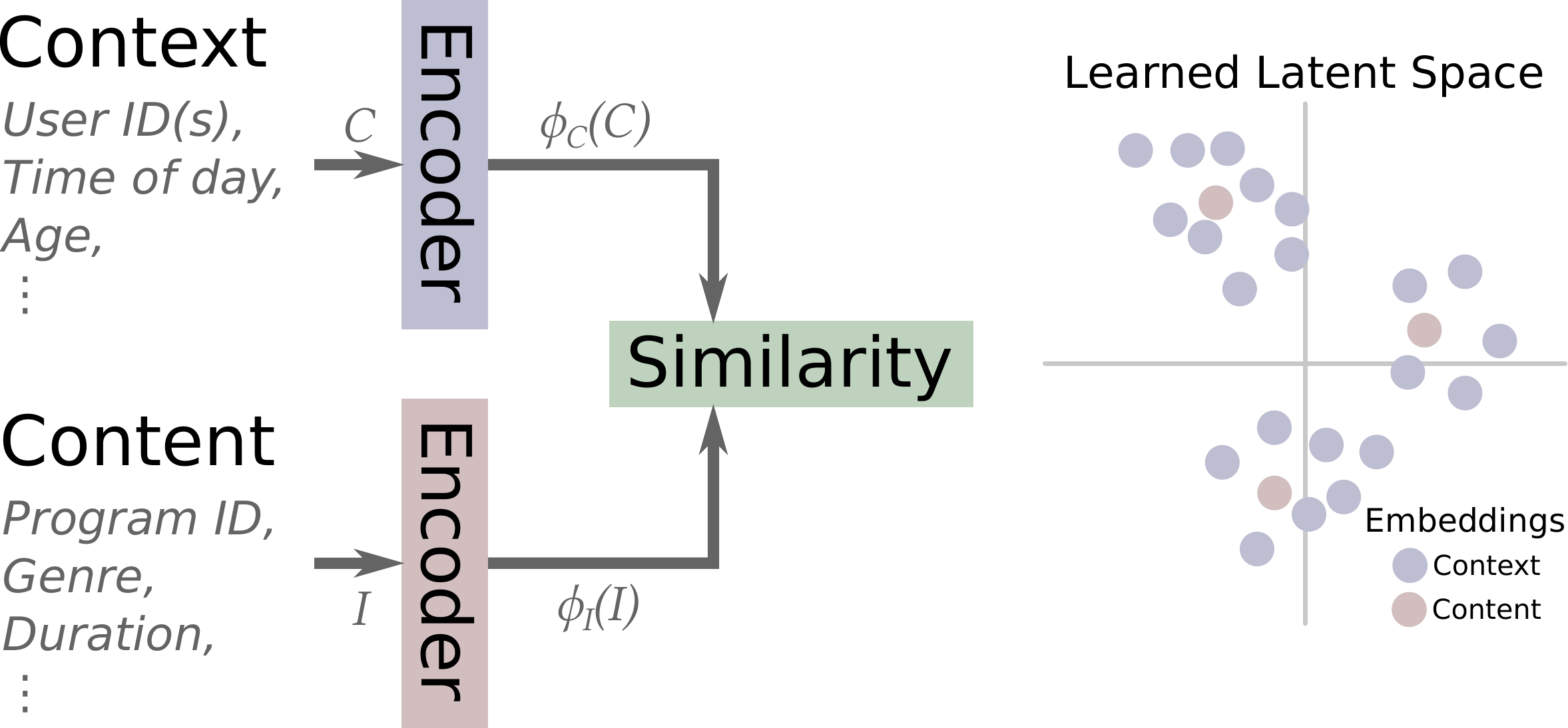}%
\caption{Framework for learning a shared latent space in order to compute similarities between context and content.}%
\label{fig:framework}%
\end{figure}%

Television exemplifies one such application in which content delivery may benefit from intelligently adapting to viewing situations, and where insights into user behavior are of high interest to stakeholders, such as content creators, schedulers, and advertisers.
Broadcasters' Audience Research Board (BARB)\footnote{\url{https://www.barb.co.uk}.} maintains a panel of households in the UK that represent the television viewing across the nation.
Each household in the panel is equipped with measuring devices, often referred to as meters, associated with each television in the home.
These meters offer a semi-automatic way for participants to report who has been watching what and when.\footnote{Presence in front of the television is registered manually using a remote control and supports multiple simultaneous users. All other variables are collected automatically.}
The data contain a large number of interactions compared to alternative data collection methods within CARS for television, such as self-reported consumption~\cite{Kristoffersen2018b}.
The large volume of reported viewing events allows data-driven learning of underlying structures as has been explored in related literature on e.g. implicit preference elicitation~\cite{Gadanho2007} and group viewing patterns~\cite{Senot2010,Chaney2014}.
It also motivates the deep learning-based methods studied in this work, where we are specifically interested in providing accurate recommendations based on contextual settings of viewing situations, while also learning representations that can help explore the complex patterns of television consumption.

To this end, we turn our attention to the framework in Fig.~\ref{fig:framework} which shows the principles behind the contributions of this paper.
The core foundation is the embedding of context and content in a shared latent space.
Here context is defined as the collection of variables specifying a viewing situation, such as user identities and temporal aspects, and content represents programs and associated metadata.
Bringing both context and content into the same latent space allows for significant advantages, such as computing similarities across domains, and ideally enables us to learn to group contexts with relevant content while also grouping contexts that show similar preferences in terms of content, and content that tend to be consumed in similar contexts.
Also, content embeddings are static at serving time and only have to be computed once (until the model is updated, e.g. with new content), which allows for efficient recommendations -- a forward pass of the context encoder and similarity computations with the precomputed content embeddings.

We propose the joint context-content embeddings (JCCE) method which learns a metric to encode contextual preferences of content.
Compared to other methods, such as Wide \& Deep~\cite{Cheng2016}, we keep context and content encoding distinct, thereby allowing efficient inference as well as independent investigations of each domain.
By using the $N$-pairs mini-batch loss \cite{Sohn2016}, we train both encoders simultaneously to optimize a joint objective.
The approach should apply to generic RS, but we focus on its application within the television domain.
A dataset consisting of 2.7M viewing events is used to demonstrate the recommender performance of JCCE.
In addition to the results obtained for recommendation, JCCE enables qualitative insights into the relationships among contextual settings and consumed content properties.
In Section~\ref{sec:rep}, we highlight this capability by displaying visualizations of embeddings, and using those we show meaningful structures in the learned latent space.

\subsubsection*{Related Work on Latent RS and CARS}
Related research in RS has investigated embeddings as a powerful tool for recommendation.
Early work for learning user and item latent representations with neural networks was proposed in~\cite{Salakhutdinov2007a}. 
Recently, the success of Word2Vec~\cite{Mikolov2013} inspired the product embeddings from purchase sequences in Prod2Vec~\cite{Grbovic2015}, which was expanded to include product side information in Meta-Prod2Vec~\cite{Vasile2016}.
\cite{Guo2016} shows competitive results in a Kaggle competition using simple mappings of categorical variables to dense vector representations.
They use concatenated entity embeddings and continuous variables as input for a neural network.
In \cite{Twardowski2016}, item and event embeddings are introduced for session-aware recommendations.
Collaborative metric learning is coined in~\cite{Hsieh2017} and demonstrates learning of a joint user-item metric.
\cite{Covington2016}~presents a real-world implementation for video content, while Wide \& Deep~\cite{Cheng2016} proves efficient for recommending apps.
DeepCross~\cite{Shan2016} use a similar framework, but they replace the standard feed-forward network with a residual network.
Neural Factorization Machines (NFM)~\cite{He2017a} combines the successful Factorization Machines (FM)~\cite{Rendle2010} with neural network architectures.
Convolutional FM (CFM)~\cite{Xin2019} uses 3D convolutional neural networks to model high-order interactions between contextual variables.

\section{Proposed Approach}
We define a viewing event as a contextual setting, e.g. temporal aspects such as the time of day, together with consumed content, e.g. genre.
Our goal is to learn joint embeddings of context and content, such that the representations of a true pair are close together, and those pairs that are unlikely to co-occur are far apart.
As an example, when a child is watching, children's content should be preferable to horror movies by having a smaller distance to the context in the latent space.
Since the data collection method relies on implicit feedback, undesired correlations between content and context are generally not present in the data.
As an example, there are no negative examples to indicate that children should not watch horror movies.
We thus have to carefully consider how we formulate the learning objective and sampling strategy.
For the remaining part of this work, we will present how JCCE is designed to overcome these challenges and achieve the goal described above.

Viewing events are logged as a pair $(I,C)$ of consumed content and contextual features, respectively.
The specific data used in this paper is described in Section~\ref{sec:setup}.
Let $\phi_I(I)$ be an embedding of the content, where $\phi_I$ is the content encoding function shown by the red block in Fig.~\ref{fig:framework}.
Furthermore, let $\phi_C(C)$ be an embedding of the context, such that the dimensionality of $\phi_I(I)$ and $\phi_C(C)$ are identical.
This allows the embeddings to exist in a shared latent space with the possibility to compute similarity scores.

\subsection{Encoder Architecture}
A contextual setting, $C$, is a sparse high-dimensional collection of numerical and categorical features describing the aspects of a viewing situation. 
These aspects include information about who is watching television as well as when and where it takes place.
Thus, one of the tasks of the context encoder is to combine the various contextual aspects of a viewing event into a single point in the latent space such that the low-dimensional representation effectively embodies crucial characteristics of the situation at hand. 

In order to embed vectors of context and content, we introduce the two encoders
$\phi_I: \mathbb{R}^{|I|} \rightarrow \mathbb{R}^{E}$ and 
$\phi_C: \mathbb{R}^{|C|} \rightarrow \mathbb{R}^{E}$, where $E$ is the dimensionality of the embeddings.
As in \cite{Vasile2016}, we set ${E=50}$ based on empirical findings, and do not investigate effects of changing $E$ further, but refer the reader to related literature, e.g.~\cite{Yin2018}.
We employ nonlinear encoder networks consisting of three fully connected layers each\footnote{Note that the two encoders are not required to have identical architectures, but in this work we have opted for similar structures.}.
The first two layers are each defined to have 250 rectified linear units (ReLUs), and the last layer is a simple linear transformation with 50 units.

\subsection{Jointly Training Context-Content Encoders}
We follow a supervised approach for learning the weights of the two encoders simultaneously.
Specifically, we use the $N$-pairs loss~\cite{Sohn2016}:
\begin{equation}
    \mathcal{L}_\text{NP}(\mathcal{X},\mathcal{Y}) = - \frac{1}{N} \sum_i^N \log \left( \frac{e^{x_i{}^\mathsf{T}y_i}}{\sum_{j}^N e^{x_i{}^\mathsf{T}y_j}} \right) +\lambda \left( \Vert x_i \Vert ^2 + \Vert y_i \Vert ^2 \right),
    \label{eq:npairs}
\end{equation}
where $N$ is the number of pairs in a mini-batch, $\lambda$ is the regularization strength, and $\mathcal{X}=\{x_1,\ldots,x_N\}$ and $\mathcal{Y}=\{y_1,\ldots,y_N\}$ are sets of $E$-dimensional vectors $x_i, y_i \in \mathbb{R}^{E}$ such that $(x_i,y_i)$ is a pair and $x_i \neq x_j$ for all $i,j = 1,\ldots ,N$ with $i \neq j$.
Thus, each $x_i$ is compared to one positive example, $y_i$, and $N-1$ negative examples, $y_{j\neq i}$.

For the purpose of training $\phi_I$ and $\phi_C$ jointly, we define $N$ to be the number of context-content pairs in each mini-batch.
The content samples follow the notation $\mathcal{I} = \{I_1,\ldots,I_N\}$. 
Likewise for context samples $\mathcal{C} = \{C_1,\ldots,C_N\}$.
Each pair $(I_i,C_i)$ is a viewing event observed in the training data, and the $N$ pairs in a mini-batch are selected such that $\mathcal{I}$ contains $N$ unique content types.
These pairs are weighted equally, but a natural extension is to consider varying confidence levels, e.g. using duration of the event. 
We define the JCCE training objective, for efficacy, as minimizing the asymmetric $N$-pairs loss of the embeddings in both directions:
\begin{equation}
    \mathcal{L}(\mathcal{I},\mathcal{C},\phi_I,\phi_C) = \mathcal{L}_\text{NP}(\phi_I(\mathcal{I}),\phi_C(\mathcal{C})) + \mathcal{L}_\text{NP}(\phi_C(\mathcal{C}),\phi_I(\mathcal{I})).
    \label{eq:loss}
\end{equation}
In addition to grouping context-content pairs, this process will also cluster 1) the contextual settings exhibiting similar preferences in terms of consumed content, and 2) the content types consumed in similar contexts.
Each of these properties can be accessed independently using the respective encoding function, and ultimately allows us to generalize the model to relationships between unseen context-content pairs.

\subsection{Recommendations}
The framework allows us to recommend content based on the contextual settings of a specific viewing situation.
That is, if we know the context, $C$, we can compute a score for some available content, $I$, according to cosine similarity $\mathcal{S}(I,C) = \phi_I(I)^\mathsf{T}\phi_C(C) / |\phi_I(I)| |\phi_C(C)|$.
The resulting recommendation is a list of all available content sorted with decreasing similarity score to the given viewing context.

\subsection{Linear JCCE}
In addition to the standard JCCE described above, we also show the performance of linear JCCE (L-JCCE); A fully context-aware configuration that uses the loss defined in Eq.~(\ref{eq:loss}) with linear encoders, $\phi_I$ and $\phi_C$.
Specifically, $\phi_I(I_i) = W_II_i+b_I$ with $W_I\in \mathbb{R}^{E \times |I_i|}$ and similarly $\phi_C(C_i) = W_CC_i+b_C$ with $W_C\in \mathbb{R}^{E \times |C_i|}$.

\section{Experiments}
In this section we conduct experiments with two main purposes; 1)~Quantitative evaluation of recommender performance of JCCE and four baseline methods; 2)~Qualitative insights through visualizations of JCCE learned embeddings.
\subsection{Setup}\label{sec:setup}
The proprietary dataset used for the experiments comprises approximately two months of television viewing within the BARB panel in the period June to July 2018.
During that time frame, 5923 households encompassing 13K unique panel members reported at least once.
We include 11 attributes of viewing events and remove all viewing with a duration of less than three minutes, since we can assume that users did not engage with the content if they watched it for less than that (also used in the official reach figures by BARB).
We also remove viewing of content with few total observations, which reduce the number of viewing events from 4M to 2.7M.
We separate the first 90\% of events into a set for training and the remaining 10\% into a set for evaluation.
The test set covers approximately one week.

Television content distinguishes itself from popular recommender domains, such as movies, in several aspects. 
Most notably, television content catalogs are time-constrained and dynamic~\cite{Turrin2014}. 
In the present contribution we focus on high-level recommendations of content genres, since these pose as robust descriptors that do not suffer to the same degree as specific programs from the rapidly changing catalog.
The reduced BARB dataset contains 94 genres from 13 top-level genres, e.g. \textit{regional} under \textit{news}.

We train the model using the Adam optimizer~\cite{Kingma2014} with early stopping, and apply dropout~\cite{Srivastava2014} to encourage contributions from more contextual variables and hence reduce overfitting decisions towards a limited selection of input features.

The evaluation is focused on the ability to recommend content (i.e. genres among the 94 available) given contextual settings.
For this purpose we report hit ratio (HR@K) and mean reciprocal rank (MRR).
That is, for each test sample a hit is achieved if the target content is within the top-K recommendations.
The resulting HR@K is the hit ratio over all test samples.
MRR is the average reciprocal placement of targets in the recommendation lists.

\subsection{Baseline Methods}
\subsubsection{Random}
A weak baseline that randomly ranks content and mainly serves as an indicator of the chances of coincidental hits.

\subsubsection{Toppop}
A context agnostic baseline that ranks content according to the number of observations in the training set.
It also serves as a measure of dominance among the most popular content compared to the less watched.

\subsubsection{Toppop (temp)}
Temporal context is a key indicator due to the strong habitual preferences in everyday television consumption~\cite{Turrin2014}. A simple, but often well-performing, baseline ranks content according to occurrences at specific temporal settings.

\subsubsection{Wide \& Deep \cite{Cheng2016}}
A state-of-the-art framework consisting of a wide component for memorization and a deep component for generalization.
We use two cross-product transformations for the wide component: 1) user IDs and genre; 2) temporal settings (day of week, time of day) and genre.
The deep component uses all features and the architecture is chosen to be similar to that of the JCCE encoders.
The model is trained using a logistic loss with observed context-content pairs as positive examples and a similar number of randomly sampled unseen pairs as negative examples.

\subsection{Results}

\begin{figure*}[!b]
    \setcounter{figure}{2}
    \centering%
    \begin{subfigure}[b]{0.32\textwidth}
        \includegraphics[width=\textwidth]{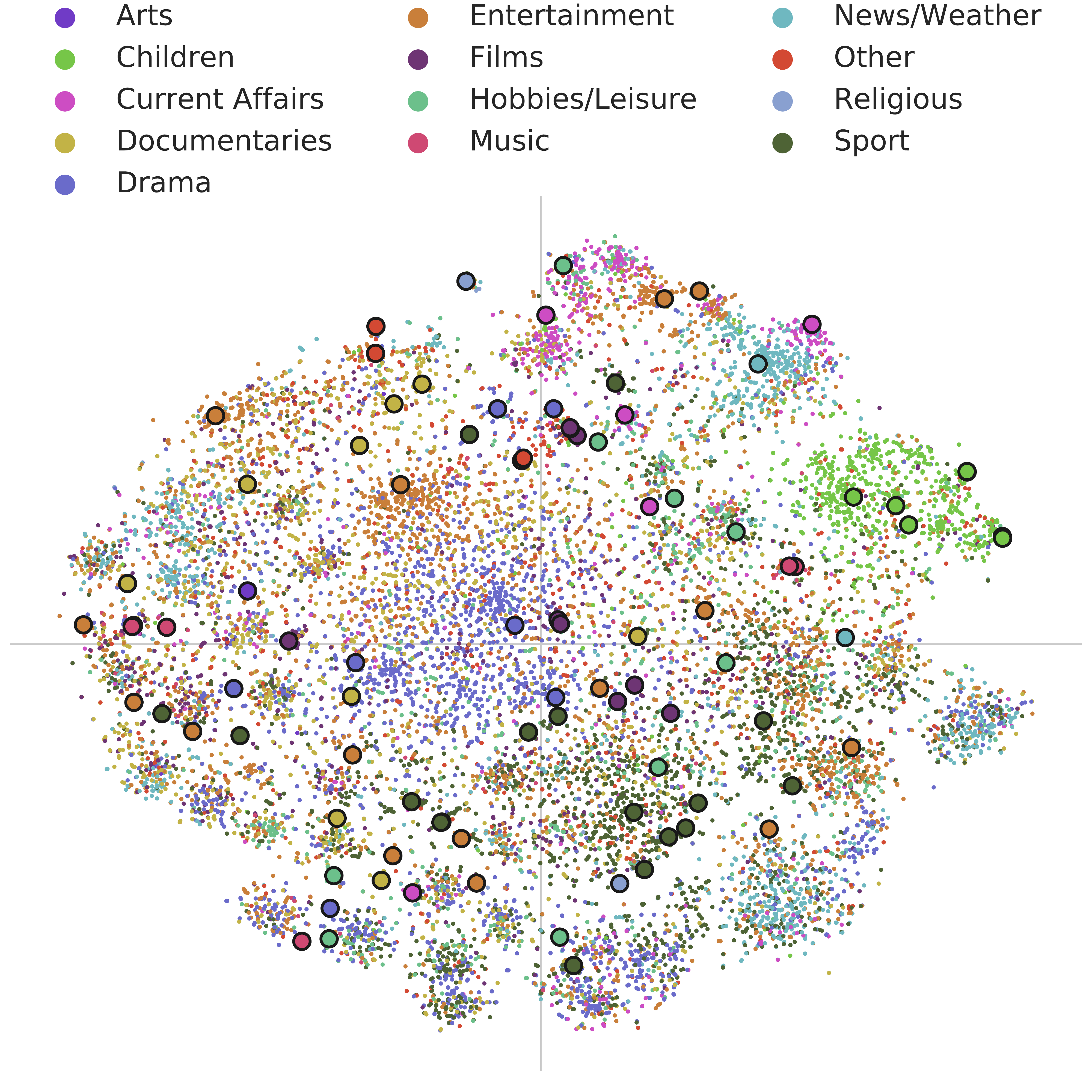}
        \caption{Observed genre}
        \label{fig:emb_a}
    \end{subfigure}%
    \begin{subfigure}[b]{0.32\textwidth}
        \includegraphics[width=\textwidth]{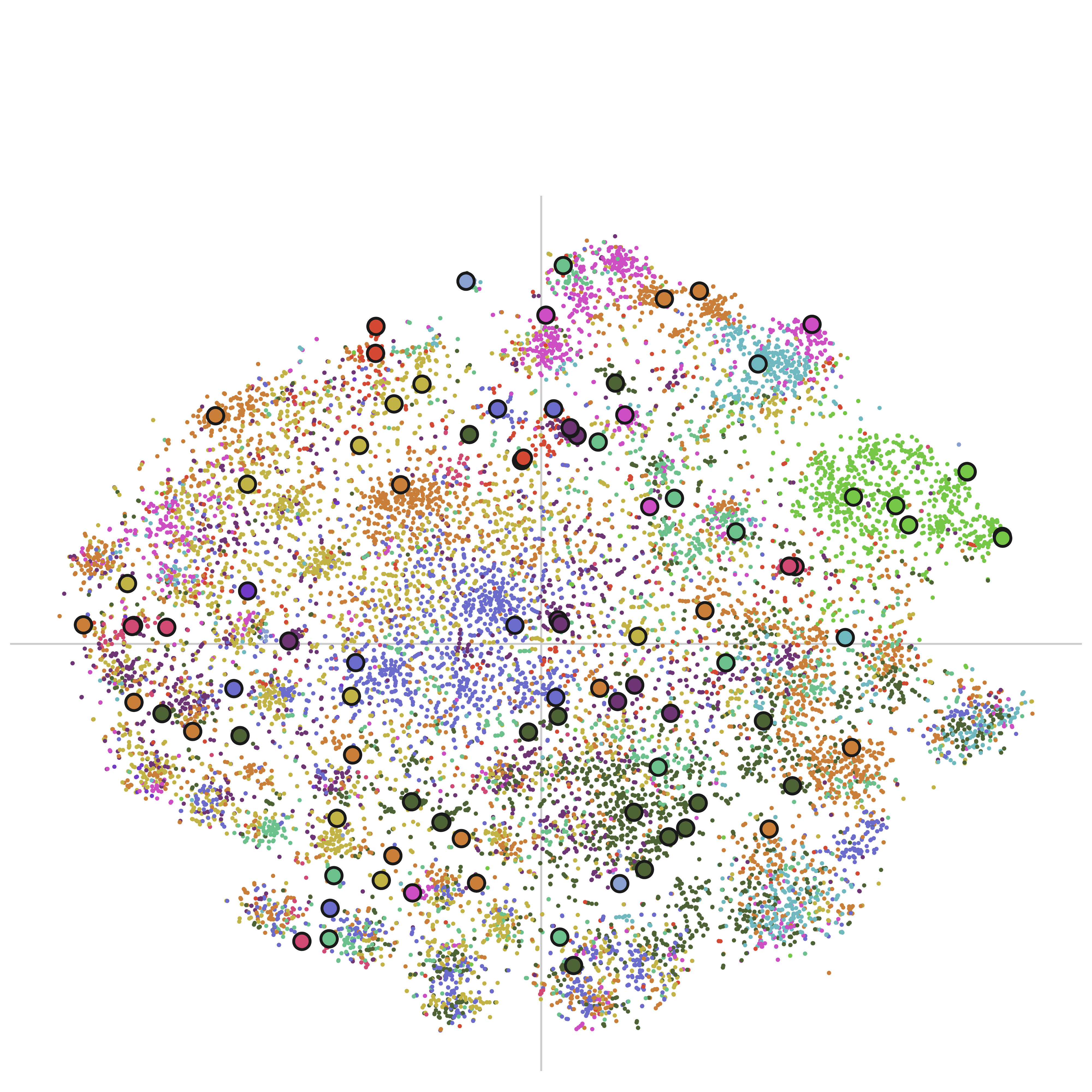}
        \caption{Recommended genre}
        \label{fig:emb_b}
    \end{subfigure}%
    \hfill%
    \definecolor{plotgrey}{RGB}{204,204,204}%
    \textcolor{plotgrey}{\vrule width0.03em}%
    \hfill%
    \begin{subfigure}[b]{0.32\textwidth}%
        \includegraphics[width=\textwidth]{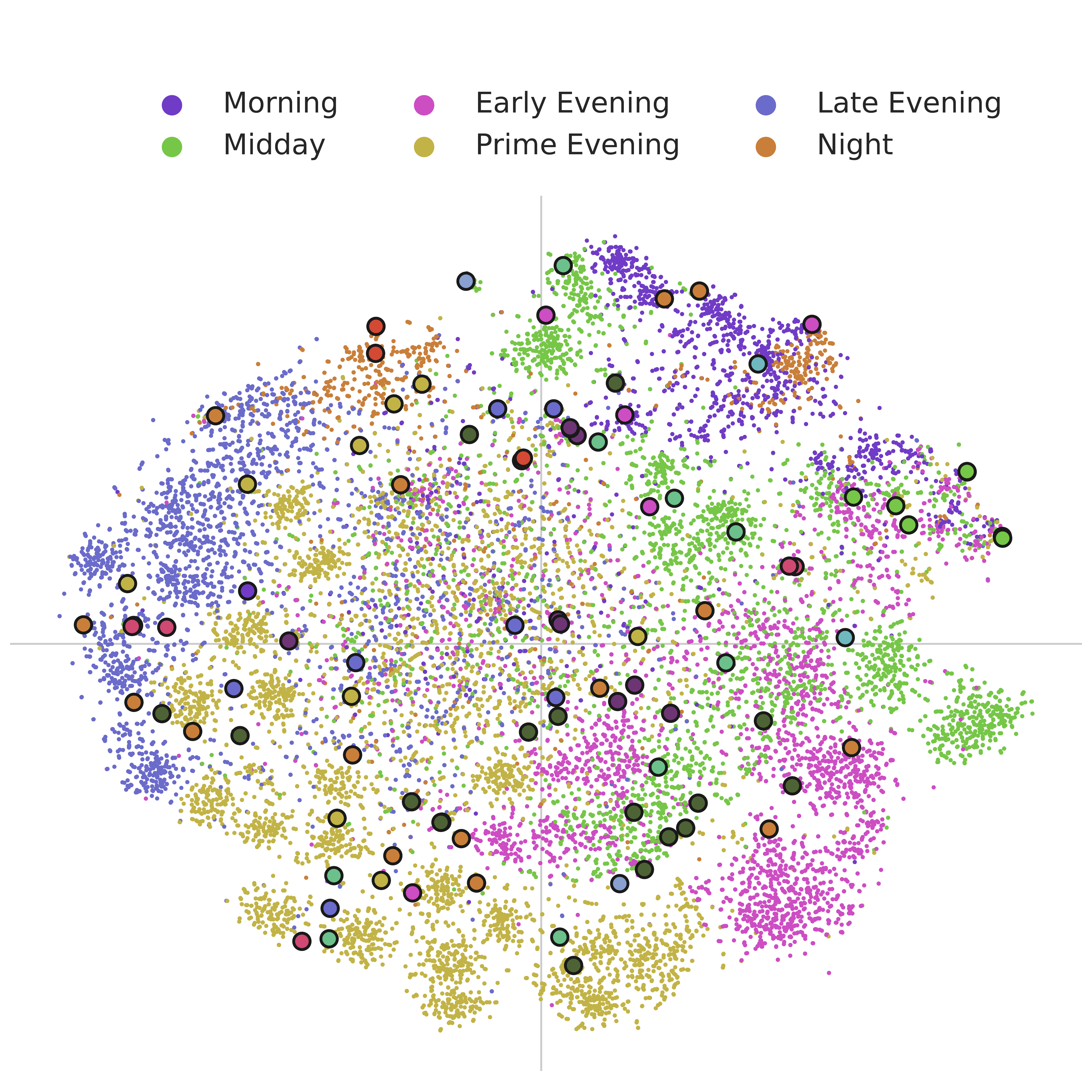}
        \caption{Time of day}
        \label{fig:emb_c}
    \end{subfigure}%
    \caption{Embeddings of the JCCE model visualized using t-SNE. The small points are context embeddings of randomly selected samples from the test set colored by (a) associated ground-truth genre, (b) genre that achieves the highest cosine similarity with the context in the 50-dimensional space, and (c) time of day. Each large point is a content embedding of a sub genre (e.g. \textit{History} is a sub genre of \textit{Documentaries}), and follows the color scheme of (a). The figures are zoomable.}
    \label{fig:content_emb}
\end{figure*}

\begin{table}[b]
\centering%
\caption{Quantitative results.}%
\label{tab:res}%
\begin{tabular}{lccc}%
\toprule%
\textbf{Method} & \textbf{HR@1} & \textbf{HR@3} & \textbf{MRR}\\
\midrule
        Random & 0.011 & 0.032 & 0.055 \\
        Toppop & 0.080 & 0.187 & 0.199 \\
 Toppop (temp) & 0.120 & 0.282 & 0.262 \\
        L-JCCE & 0.222 & 0.434 & 0.376 \\
  Wide \& Deep & 0.264 & 0.467 & 0.409 \\
          JCCE & 0.293 & 0.506 & 0.443 \\
\bottomrule%
\end{tabular}%
\end{table}%

Table~\ref{tab:res} compares performances of the methods.
Firstly, note the hit ratio scores of Toppop suggesting that the most seen genre in the training set accounts for 8\% in the test set, while the three most popular genres make up a total of 18.7\% of the total viewing.
There is a notable performance gain when taking temporal aspects into account in Toppop.
It is also evident that L-JCCE is less accurate than its nonlinear counterpart, most likely due to a lower capacity both in terms of the number of parameters and the inability to model nonlinear interactions.
JCCE outperforms Wide \& Deep with relative improvements of 11\%, 8.4\%, and 8.3\% for HR@1, HR@3, and MRR, respectively. 
The difference is partially explained by the training procedure, where JCCE utilizes the $N$-pairs strategy, while Wide \& Deep relies on the point-wise logistic loss.

\begin{figure}[t]
    \setcounter{figure}{1}
    \centering%
    \includegraphics[width=\columnwidth]{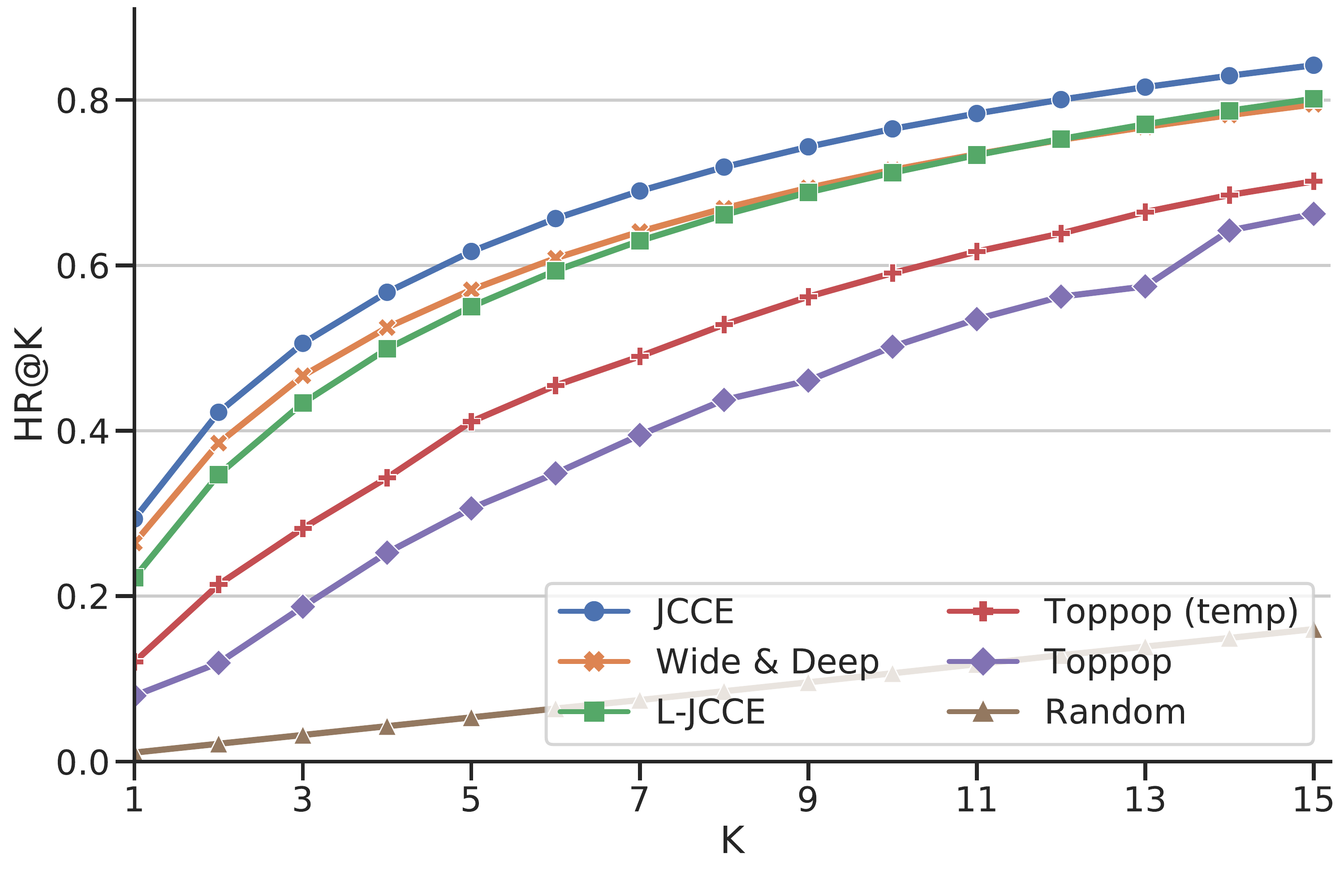}%
    \vspace{-1.5em}
    \caption{Hit ratio performance at increasing K.}%
    \label{fig:HRatK}%
    \setcounter{figure}{3}%
    \vspace{-1em}
\end{figure}

Fig.~\ref{fig:HRatK} shows how the methods perform in terms of hit ratio for different settings of K.
As can be seen, JCCE shows consistent improvements compared to the other methods, which has been further inspected by conducting McNemar's tests, verifying that all improvements are statistically significant with $p<0.001$.

\subsection{Analysis of Learned Representations}\label{sec:rep}
In this section, we take a qualitative look at how contextual settings and content group in the learned latent space. 
To this end, a large set of viewing events are randomly sampled from the test set, such that they are not seen during the training phase.
We then compute their embeddings using the context encoder, $\phi_C$.
In Fig.~\ref{fig:content_emb} we have used JCCE and reduced the embeddings from 50 to 2 dimensions for visualization with t-SNE~\cite{Maaten2008} using cosine similarity.
We also include content embeddings from each unique genre using $\phi_I$.

The three plots of Fig.~\ref{fig:content_emb} show the same embeddings colored according to three different variables of the viewing events.
The first one, Fig.~\ref{fig:emb_a}, colors the context embeddings from the associated target content genre.
Note e.g. the cluster of contexts with children's content, and how news/weather and current affairs group together in the top-right, suggesting that they tend to be consumed in similar contextual settings.
Furthermore, the content embeddings of these genres are located in the same areas.
The second plot is related to the first, but instead of showing ground-truth genre choice it displays recommended genre by JCCE, and serves as a qualitative tool to assess the performance and types of errors made by the model.
The~\ref{fig:emb_b}~plot of ideal RS would thus be similar to the~\ref{fig:emb_a}~plot.
The last plot, Fig.~\ref{fig:emb_c}, clearly demonstrates the strong temporal influence on the learned representations.
Combined with more contextual variables, e.g. social, these embedding visualizations enable valuable insights, whether investigating patterns of context or content.

\section{Conclusion}
In this work, we explored deep embeddings learned jointly for context and content.
We introduced the unified framework of JCCE that delivers context-aware recommendations, while also supplying tools for exploring patterns of context-content, context-context, and content-content relationships.
We demonstrated the capability of JCCE by achieving superior performance for recommendations in the television domain, and inspected the interpretability by visualizing learned structures in the shared latent space.
For future work we will explore the shared latent space, as well as evaluate the approach against more baseline methods on publicly available datasets suitable for CARS.

\balance

\end{document}